-------------------------- Plain LATEX file --------------------------
\documentstyle[12pt]{article}
\newcommand\be{\begin{eqnarray}}
\newcommand\ee{\end{eqnarray}}
\begin{document}

\begin{tabbing}
\`SUNY-NTG-93-24\\
\`Nov. 1992
\end{tabbing}
\vbox to  0.8in{}
%double spacing:
%\vbox to  0.4in{}
\centerline{\Large \bf Gluon multiplication}
\centerline{\Large \bf in high energy heavy ion collisions}
\vskip 2.5cm
%double spacing:
%\vskip 1.25cm
\centerline{\large  L.Xiong  and E.Shuryak }
\vskip .3cm
\centerline{Department of Physics}
\centerline{State University of New York at
Stony Brook}
\centerline{Stony Brook, New York 11794}
\vskip 0.35in
\centerline{\bf Abstract}
\indent
Hot gluons are the dominant components of the QCD plasma to be formed in
future high energy heavy ion experiments. In this paper we study the elementary
processes
in the plasma medium
for gluon multiplication based on all orders of
the tree-diagrams in perturbative QCD.
When applying to the
chemical equilibration in the expanding system, we found that the
gluons reaches chemical equilibrium well within its plasma phase.
The inclusion of all the next-to-leading order processes makes the
equilibration considerably faster than the simple $gg\leftrightarrow ggg$
one considered previously.

\vskip .25cm
\vfil
\noindent
${^\dagger}$Supported in part by the US Department
 of Energy under Grant No. DE-FG02-88ER40388.
\eject

\newpage
\pagestyle{plain}
\setlength{\baselineskip}{22pt}
%\addtocounter{page}{-1}
\vfill\eject
\centerline{\bf 1. Introduction}
\medskip

In the upcoming relativistic heavy-ion collision experiments
at RHIC and LHC, one is hoping  to produce  the
{\it quark-gluon plasma}, a deconfined and chirally symmetric
phase of QCD,
during the first several fm/c of the collisions.
Some basic questions regarding the plasma one would like to discuss
include the followings:
What is
the 'equilibration time' $\tau_0$ from which the momentum distribution
is equilibrated?
What is the 'initial temperature' value $T_i$ at this moment?
Is it well above the critical point $T_c$, so that one can use the
 perturbation theory?
With which accuracy this concept
makes sense, say how accurate is
a thermodynamic relations between energy and entropy densities?
 What is the composition of the matter at this moment?
What are the most unambigious signals, which provide experimental
estimates of these parameters?

As early as in \cite{Shuryak_80},
some key elements of the qualitative features of the plasma were already
proposed. Those included very small gluon mean free paths and
predictions of relatively high initial temperatures.
Later studies have confirmed that
at RHIC and LHC energies one seems to
 enter a new dynamical regime, characterized
by the so called {\it 'semi-hard'}
processes, which involve partons with momenta (and momenta transfer)
 $p\sim$ 1-3 GeV (known also as
'mini-jets'). Their role in nuclear collisions was discussed by
\cite{Kajantie_minijets1,Kajantie_minijets2,Eskola_etal,Blaizot_Mueller,Wang_92}
. It is realized up to now that such parton collisions
can no longer be considered as
isolated rare events, but are involved in many complicated,
cascade-type processes. Perturbative QCD can and should be used in order to
answer these questions.
At the same time, any color field should rapidly disappear
in such plasma, so
e.g. the string-based models can no longer be used.

  The most significant changes in common expectations
 on the physical conditions of the
plasma at the early stages have taken place during the last  year.
The so called 'hot glue scenario'
\cite{Shuryak_twostage} is based on perturbative QCD and it
 predicts the initial temperature $T_i \approx 500 MeV$ \footnote{Here and
below
we give for orientation some approximate numbers, corresponding to
central AuAu collision at RHIC, 100+100 GeV/nucleon.}. This is
about twice higher, than considered previously in
the so called 'standard scenario' (see e.g. \cite{Satz_QM91}) which
assumes {\it complete} equilibration of QGP
 by the time $\tau_0= 1 fm/c$ and leads to the
initial temperatures
$ T_i \approx 240\, MeV (RHIC), T_i \approx 290\, MeV (LHC)$.
Clearly, such  dramatic difference affects all predictions, for example
for charm
enhancement \cite{Shuryak_twostage},
 dileptons and photons
\cite{Shuryak_Xiong}, etc.

Also during the last year
the first results of the quite ambitious program, the 'partonic cascade model'
(PCM) \cite{Geiger_Muller}, have been published. This model
aims to trace the partonic system evolution
all the way, from the
structure functions  to hadronization.
Generally speaking, these results
\cite{Geiger_thermo,Geiger_particle}
 strongly support the 'hot glue' scenario.
However, there is some difference in the method used by PCM and this work,
and we would like to comment on it.

A major approximation of PCM is the general idea of
{\it cascading of  virtual gluons}, in which new ones are
produced in a sequence of processes
 $g^* \rightarrow g^*g^*$, where star means positive invariant mass.
However, in general a virtual gluon is a gauge non-invariant concept, and
therefore this approach  can only be used under specific assumptions. In PCM
the Lipatov-Altarelli-Parisi  branching functions are applied, which is
only possible for {\it small-angle soft-gluon
radiation} and only  in the {\it leading log} approximation.
 First of all, soft radiation is exactly
the process which should be strongly affected by the plasma screening effects.
Large-angle non-soft radiation is to be looked at with greater attention, and
this is what we are going to do in this work.

  Our second general question is as follows.
 Even speaking about soft radiation, one
may question  a concept of cascading, with
probablities rather than amplitudes. In deep inelastic lepton-hadron
scattering one
 can justify it by the fact, that   in the {\it leading log} approximation.
partons form a natural sequence of virtualities, $Q^2 \gg q^2_1 ... \gg q^2_n$,
and therefore no interference is possible, since each parton is in its own
kinematical domain. However, in general this is not the case
for the gluon multiplication
process to be considered.

   Clearly, one has to consider the multi-parton processes in wider context,
separating all 'short-time' processes, for which a cascade approximation is
justified, from the 'long-time' ones, such as interaction with collective
soft modes. Parametrically speaking, at high T
and small g(T) one should separate 'long' time scale
of binary gluon collisions ( $t_{long}\sim O(1/g^2 T)$
if small angle scatterings are relevant or $t_{long}\sim O(1/g^4 T)$
for large angles) from  the 'short-time' one, for which we take the
'electric screening scale' $t_{short}\sim O(1/g T)$\footnote{
Available lattice data suggets that for plasma under consideration, with
$T\sim 2-3 T_c$ the magnetic screening length is numerically {\it smaller}
than electric one. However, parametrically speaking, at very large T
it is $O(1/g^2 T)$, and therefore it is undistiquishable
from the distance between small-angle scattering events.}.
All multi-parton processes should be included and properly cut off,
at 'short' time scale. The virtuality of external lines is of the order
$p^2 \sim 1/t^2_{long}$, and ignoring them\footnote{
Note that this approximation has  {\it power} accuracy in terms of the
ratio of two scales. For soft radiation this leads to additional suppression,
however, known as Landau-Pomeranchuck-Migdal effect.}
one deals with  the matrix elements
of the process {\it on the mass shell},
a manifest gauge invariant quantity. This is the only consistent way
of treating the cascade.

There are also some remaining problems, which are
not specific to PCM. Those include: (i) rather uncertain 'initial
conditions': e.g.
possible correlations between parton momenta
and positions, which can significantly
modify the results; (ii) poor
accuracy of the general treatment of the higher order
parton multiplication processes; (iii) the in-matter infrared cut off, which
is not yet  implemented in a selfconsistent way.

   In the present paper we address the second problem, that of parton
multiplication. It was also discussed
in a recent paper \cite{Biro_etal} the simplest relevant process
$gg \rightarrow ggg$ is studied under several approximations
(see below).
The main conclusion of this paper is that
gluon chemical equilibration takes too long to happen in realistic collisions.
However, this is partly due to the small number of initial gluons
assumed, following HIJING predictions.
The rate of gluon multiplication due to the same process
$gg \rightarrow ggg$ was also discussed in
\cite{Lichard_Prakash}, using the exact matrix element \cite{Berends_2to3}.

   This paper is organized as follows. Our model for the plasma and
the particle multiplication is discussed in section 2;
in section 3 we present a detailed analysis of
$gg \rightarrow ggg$  reaction rate; and will proceed to higher order
 gluon multiplication processes in section 4;
Finally, in section 5 we solve the equations for
gluonic temperature and fugacity, and derive conclusions
on
chemical equilibration in the
expanding gluonic plasma, which are summarized in section 6.

\vskip 1cm
\centerline{\bf 2. The model}
\vskip 1cm
\noindent{2.1 Kinetic and chemical equilibration of gluons }
\vskip 1cm
   In contrast to such global approaches as PCM, we only discuss
some particular stage of the time evolution of the system, and therefore
we can significantly simplify the problem.

  First of all,
in this work we concentrate on the
gluon components only. This is due to the following two reasons.
The first is that
the gluons dominate
the nucleon structure functions at small x: their number is about twice
larger than that for
the sea quarks and antiquarks of all flavors:
$ N_g/N_{q+\bar q} \approx 2$. (This can be compared to
the one in equilibrated quark-gluon  plasma $N_g/N_{q+\bar q} \approx 1/2$ ).
The lowest order process  $gg\rightarrow q\bar q$,
has a rate about
 two orders of magnitude smaller than gluon elastic
scattering, so it cannot balance the ratio.

  The other reason is that gluons rescatter very actively. The
cross section of the $gg$ elastic scattering has
the matrix element being \cite{Combridge_etal}
\be
|{\cal M}_{gg\rightarrow gg}|^2 =
{9 \over 2}(3-{ut \over s^2} -{us \over t^2}
-{st \over u^2}) \label{eq:2to2}
\ee
which is is very large, both for small angle scattering (relevant for color
changing) and large angle ones (relevant for equilibration of momenta).
Therefore, the gluon subsystem reaches {\it kinetic} equilibration of momenta
rapidly
\cite{Shuryak_80,Shuryak_twostage}, at  $\tau_0 \sim 0.3$ fm/c.

  This is our second approximation:
 from this time  on, we assume kinetic equilibration of gluon momenta
\footnote{The quark subsystem reaches kinetic equilibrium with the gluons
at almost the same time because of the big $qg$ elastic cross section. The
quark momentum distribution can be parametrized in a similar fashion,
and similar equation for quark fugacity shows that is grows too slowly.}.
We use a simple
parameterization for gluon
 distribution function\footnote{We use Boltzmann form just for simplicity
of the equations to follow, Bose parameterization only differs at small momenta
which are cut off anyway.}
, which
contains two time dependent parameters, 'fugacity' $\xi$ and temperature $T$
\be
 f(p, \tau) = \xi(\tau) \cdot  exp(-p/T(\tau)). \label{eq:mom}
\ee

The {\it chemical} equilibrium of the system implies  vanishing of the
chemical potential or $\xi=1$.
Such equilibration goes via creation and annihilation of gluons
rather than rescatterings, and
it is the objective of this paper to evaluate their rate and to see
how quickly these processes lead the
system toward chemical equilibrium.

\vskip 1cm
\noindent{2.2  The initial conditions }
\vskip 1cm

Obviously the next important question is
about the {\it initial conditions},
the values of $T_i$ and $\xi_i$ at the onset of kinetic equilibration,
time $\tau_0$.

Although now it is generally agreed that the kinetic equilibration time
is short, $\tau_0 \sim 0.3$ fm/c,
the value of the initial fugacity $\xi_i$ remains very uncertain.
For example, PCM optimistically
suggests $\xi_i \approx 1$, indicating that the gluon system would even
reach chemical equilibrium from above;
while estimates based on HIJING yields only $\xi_i \approx 0.1$, probably
the possible lower bound.

  Although we do not discuss the issue of these differences in this paper,
by keeping the initial condition as adjustable input, we can still
 provide some guidelines for it.
In principle, one can think about the initial condition problem
in  two ways, either going
(i) {\it forward} in time, starting from the
structure function and follow the parton cascade evolution; or
(ii) going {\it backward in time} and  looking for the
time moment at which the scattering and expansion rate become equal
\cite{Shuryak_twostage}.

  The first way is difficult and depends on many unknown things, like the
spacial correlations between partons and interference of the medium.
The only thing we know is the initial number of the partons inside
the system, from the structure functions.
We need partons carrying few percent of the total momentum, and those
are gluons and   'sea' quarks (and antiquarks).
One can see that if adding them up, the result is
\be
dN_{q+\bar q}/dy =2\int_{x_{min}}dx( u_{sea}(x)+d_{sea}(x)
+ s_{sea}(x) )
 \approx 1.\\
dN_g/dy \approx 2dN_{q+\bar q}/dy \approx 2.  \\
\ee
These are the partons pre-existing in the system.
If all of them are rescattered and thermalized, one gets for central
AuAu collisions at RHIC ${dN\over dy}^{AuAu} \sim  400$ quarks and
${dN\over dy}^{AuAu} \sim 800$ gluons.

The second approach has to
normalize the total parton multiplicity to the total
 entropy, later
observed as multiplicity of secondaries, assuming its conservation at later
stages (or correcting for its additional production on the way).
The numbers we use for numerical examples are normalized in this way  to some
prefixed standard multiplicity,  $dN_\pi/dy=1400$
for AuAu central collisions
at RHIC \cite{Satz_QM91}.

\vskip 1cm
\noindent{2.3  Reaction rates in the gluonic  plasma }
\vskip 1cm

   We define n to be the {\it total} number of gluons participating in the
process (for example $gg\rightarrow (n-2)g$), and
the rate as the number of events
per $d^4x$
\def\ph #1 { {d^3p_{#1}\over (2\pi)^3 2E_{#1} } }
\def\f #1 {{f(p_{#1}) }}
\be
R_{gg\rightarrow (n-2)g} &=& {1\over 2!\cdot (n-2)! }
\int \ph 1 \f 1 \ph 2 \f 2 \nonumber \\
& & \prod_{i=3}^n [1+f(p_i)] \ph i
| {\cal M}_n |^2
(2\pi)^4 \delta^4 ( p_1+p_2- \sum_{i=3}^n p_i ),  \label{eq:rate}
\ee
where $f(p)$ is the momentum distribution of the gluons.
  In principle, one can use it for incoming gluons at pre-equilibrium stage,
in which case f(p) is proportional to nuclear structure functions.
  In the equilibrated stage as we discuss below, we use just Boltzmann
distribution Eq.(2).

In order to apply perturbative QCD all momenta and momenta transfer
should be larger than some cut off value $q^2 > s_0$.
In vacuum it is determined by some non-perturbative phenomena, leading
to $s_0\sim 1-3 \, GeV^2$. In dense plasma of partons under consideration those
non-perturbative phenomena are believed to be absent\footnote{
It was shown by many lattice
calculations, that at $T\sim 2-3 T_c$ we discuss,
 the thermodynamical quantities and
correlators are reasonably well reproduced by the perturbative expressions.}
 and the cut off should  instead be
 determined by many-body phenomena like
Debye screening.

The lowest
order perturbative result \cite{Shuryak_plasmath} for the pure gluon plasma
leads to the Debye mass
\be m_D^2=(N_c/3)(gT)^2 \ee
{}From SU(2) and SU(3) lattice calculation \cite{Gao_screen,Irback_etal}
the Debye-screening mass was found to be
stabilized at $2m_D= 2.1 T$ in pure glue plasma for $T >
2-3  T_c$. This number also gives the 'effective charge' value.
In the rest of our study, we will use this result for the low energy cut off
\footnote{Note that in a chemical non-equilibrated plasma, the
 screening mass  should be different, $O(\xi^2)$. It increases the rates,
if implemented, but then we have to set an additional cut off valid at
$\xi=0$ (no plasma at all). We have ignored this point in the present work.}

We will require the binary invariant of each pair  ( $s_{ij}=(p_i + p_j)^2$ )
of 4-momenta $p_i$ and
$p_j$ be larger than  a constant factor times $T^2$
\be s_{ij} \ge s_0 = \eta^2 T^2. \ee
 If we neglect the temperature dependence of $g_s$,
the rate scales with temperature as follows
\be
R_{gg\rightarrow (n-2) gg} \equiv \alpha_{n}(\eta) T^4. \label{eq:alph} \ee
where the rate coefficients $\alpha_i$'s are dimensionless numbers
 depending on  the details of matrix elements and
the plasma screening parameter $\eta$ (an 'effective charge', if one wishes).

Another way of measuring the reaction rate is the
number of collisions per time per particle. The relation between the two
rates is
$\nu = dN/d\tau = n_i  R_{gg\rightarrow ggg}/\rho ,  $
where $\rho$ is the density of the particle and $n_i$ is the number of
identical particle in the initial state.
Note that $\nu$ scales as $T$.

\vskip 1cm
\centerline{ \bf 3. The simplest gluon multiplication process $gg\rightarrow
ggg$ }
\vskip 1cm

The matrix element ( summed over all the final states and averaged over
the initial state ) of the simplest process which produces an extra gluon
has been calculated in \cite{Berends_2to3} and elegantly presented as
\be\displaystyle
|{\cal M}_{gg\rightarrow ggg}|^2 = {27\over 160} \
{ \sum_{m<n} s_{mn}^4 \over \prod_{m<n}s_{mn}}
\sum_{\rm non-cyc\ perms} (12345), \label{eq:2to3}
\ee
where all gluons are on-shell, $p_i^2=0$. Here the binary invariants are
$s_{mn}= p_m\cdot p_n$ with $p_i,\ (i=1...5)$ the four momentum
of the gluons, and the so called 'string' of momenta is a shorthand notation
for a product of binary invariants
$(12345)= s_{12}s_{23}s_{34}s_{45}s_{51}$.
The formula has very symmetric appearence and each $s_{mn}$ appears
in denominator only once!

   If not cut off, the kinematic region of small $s_{mn}$
contribute dominantly to the process. This corresponds to either
 soft gluons or to two  collinear ones. In this situation,
the matrix element for three gluon final state in Eq.(\ref{eq:2to3})
can be factorized  to that for two ( Eq.(\ref{eq:2to2}) )
times a 'bremsstrahlung factor':
\be
d\sigma_{gg\rightarrow ggg} = d\sigma_{gg\rightarrow gg}
{\alpha_s N_c\over \pi} {dp \over p}{d\cos\theta\over 1-\cos\theta }
\label{eq:soft}
\ee
with $p$ the momentum of the soft gluon and $\theta$ the radiation angle.
Here one  clearly sees
both the ``soft" and the ``parallel" infrared logs, common to
the bremstrahlung processes.

The above approximation for the matrix element
(originally derived in \cite{Gunion_Bertsch} for gluon radiation from
quark-quark scattering) was used explicitly in \cite{Biro_etal}
in the form
\be
{ d\sigma_{gg\rightarrow ggg}\over d^2q_Tdy d^2k_T } \approx
{ d\sigma_{gg\rightarrow gg} \over d^2q_T }
{ N_c \alpha_s\over \pi^2 } {q_T^2 \over k_T^2(k_T-q_T)^2 } .
\label{eq:2to3a}
\ee
  However, if  the added gluon is much
softer than others, it would be cut off by
the requirements discussed in the previous section. In fact
the log of the highest to the lowest allowed momentum transfer
is not a large parameter in our case, it is about only 2. to 3.
Therefore  the leading log approximation is not reliable and one
should carry out a detailed study of large-angle non-soft radiation as well.

We have performed numerical Monte-Carlo integration for calculating the
reaction rates defined in Eq.(\ref{eq:rate}).
We have used a constant $\alpha_s = 0.3$ here and
through the paper. With the chosen cut-off from the lattice result, we get
\footnote{The  'stimulated emission' factors (1+f) are non-negligible, they
increase the rate by roughly a factor of 1.-2. here and 2.-3. later for large
n processes.}
\be
\alpha_4(\eta=1.1)=  2.65,              \ \ \
\alpha_5(\eta=1.1)=  2.32. \label{eq:alpha45}
\ee
  Thus, the gluon multiplication rate is comparable to
rescattering one. Moreover,
the $gg\rightarrow gg$ rate was evaluated
 above including the small angle scattering,
which does not change parton distributions very much.
It is well known that if
one evaluate the 'transport cross section', the
corresponding rate is smaller, $O(g^4 log(1/g)$
at small g. The process considered $gg\rightarrow ggg$
has therefore a rate
$O(g^4 log^2(1/g)$, and it is large numerically even outside of the leading log
approximation.
Those facts tell us that $gg\rightarrow ggg$ process is
probably very important part of kinetic equilibration as well, and it should be
included in discussion of viscosity and similar phenomena as well.

    Unfortunately, the leading logarithmic term due to soft gluon
approximation is not sufficiently reliable,
while other terms depend also on the cut off, or on
the value of $\eta$ parameter. Therefore
we investigated the issue somewhat
further by studying the detailed dependence of the rate on different cut off.
 For elastic collisions including small angles
it is $\alpha_4 \sim 1/\eta^2$
because of power divergence of Rutherford-like scattering.
The  gluon production process should,  as explained earlier,
have an additional double log dependence  expected.
We have studied the dependence numerically, in
the region $0.05<\eta<2.5$, and present the results as solid curves in Fig.1.
They  can be  parameterized as
\be
\alpha_4(\eta) &=& {2.6\over \eta^2}, \\
\alpha_5(\eta) &=& {0.65\over \eta^2} \log^2({ 9.0\over\eta^2}) .
\ee
In the plot the parameterization appear as the dashed curves
and they actually fit the solid curves quite well.
 It is easy to understand why
 $\alpha_4$ and $\alpha_5$
cross each other: it is because the
double log factor becomes more significant at smaller $\eta$. Physically
it means that when the resolution for gluons is small, gluons would
rather radiate easily.
The cross over point is at about $\eta=1$, which corresponds almost
the same as the
screen mass coming from the lattice studies.

Our results can be compared with some other recent works. In
our notations, those of ref. \cite{Lichard_Prakash} are
(based on the same exact matrix elements in Eq.(\ref{eq:2to3})
\footnote{
The symmetrization factors $1/2!$ and $1/3!$ for the final state gluons
are  missing in the results of \cite{Lichard_Prakash}
for $gg\rightarrow gg$ and $gg\rightarrow ggg$ respectively, are included
here. Also we have rescaled their results due to the different $\alpha_s$ being
used. }
\be
\alpha_4(\eta=1.58)= 0.60  \ \ \
\alpha_5(\eta=1.58)= 0.20 \ \ \  \ee
which are a little bit different than ours
\be
\alpha_4(\eta=1.58)= 0.90  \ \ \
\alpha_5(\eta=1.58)= 0.30 \ \ \  \ee
We notice that
the ratios of the two rates  are identical though.

The result of ref.
 \cite{Biro_etal} reads
\be
\alpha_5(\eta=2.0)= 0.30  \ee
which should be compared with ours which is
\be
\alpha_5(\eta=2.0)= 0.10.  \ee
The discrepancy is quite large, a factor of 3.
\vskip 1cm
\centerline{ \bf 4. Higher order gluon multiplication processes $gg\rightarrow
(n-2)g$ }
\vskip 1cm

   One of the most interesting development in perturbative QCD is the
derivation
of the exact expression, summing contribution of all diagrams to n-gluon
processes  for
the {\it maximum helicity violation amplitude}. This is known as
the ``Parke-Taylor formula'' \cite{Parke_Taylor}
\be
 |M^{PT}_n|^2 = g_s^{2n-4} {N_c^{n-2}\over  N_c^2-1} \sum_{i>j}s^4_{ij}
\sum_{\rm P} {1 \over s_{12}s_{23}... s_{n1}}  \label{eq:pt}
\ee
In the above $s_{ij}= (p_i+p_j)^2$, the summation P is over
the $(n-1)!/2$ non-cyclic permutation of $(1...n)$.
It looks like direct generalization of n=5 case discussed above.

   Unfortunately, the
exact result for other chiral amplitudes remains unknown.
However,  assuming that they are of
the same  magnitude  as the ``Parke-Taylor'' one, one gets some estimate
for the n-gluon matrix element. This was proposed first by
Kunszt and Stirling \cite{Kunszt_Stirling} who add the following
 factor in front
of the ``Parke-Taylor'' formula
\be
 |M^{KS}_n|^2= KS(n) |M_{PT}|^2, {\rm \ \ with\ \ }
KS(n) ={2^n-2(n+1) \over n(n-1) } \label{eq:ks}
\ee

We have checked Eq.(\ref{eq:ks}) against the exact results for
$n=4$ and $n=5$ in Eqs.(\ref{eq:2to2}), (\ref{eq:2to3}), and
found that in these cases
one indeed needs the KS correction ( $KS(4)=1, KS(5)=2$ )
to recover the analytical results correctly.
For higher orders
 a number of authors \cite{Maxwell_89,Berends_etal,Kleiss_kuijf}
have checked this expression
up to $n=10$ using the Monte-Carlo generators, evaluating diagrams
directly.
They have found that Eq.(\ref{eq:ks}) does a very reasonable job, although it
consistently overpredicts the cross section slightly.
The true matrix element for $n\ge 5$ should therefore be within
the range
\be
 2  |M^{PT}_n|^2 \le  |M_n|^2 \le  |M^{KS}_n|^2 .
\ee

  Evaluation of the total cross section is a matter
 of integration over the many-body
phase space, which is analytically difficult.
 However, if the ratio of the collision energy to the cut off $s/s_0$
is treated as a large parameter, one can find its asymptotic behaviour.
Since each binary invariant happen to be present in denominator only {\bf
once,}
it is not hard to figure out  that
the leading term of the total cross section should have the double log
behavior $ \sigma_n \sim [log^2(s/s_0)]^{n-4} $.

  A simple way of observing it is related with the soft-gluon case. Suppose the
last gluon in the string is {\it softer than all others}
 (this can always be done
by changing the numeration). Then
the Parke-Taylor matrix element can be factorized as
\be
 |{\cal M}^{PT}_n|^2  \approx
( n-1 ) g_s^2 N_c
{1\over p_n^2 ( 1- \cos \theta ) } |{\cal M}^{PT}_{n-1}|^2 ,
\ee
where $p_n$ is the three momentum of the n-th gluon, $\theta $ is its
orientation with respect to any  one among the $n-1$ gluons.
The total cross section then has the form
\be
\sigma_n( \sqrt s )  \approx
 { n-1\over n-2}  {\alpha_s N_c \over 4\pi }
\int_{s_0} ^{s-s_0} {dM^2\over s-M^2} \sigma_{n-1} (M)
\int { d\cos\theta\over 1-\cos\theta }
\ee
When one proceeds iteratively, it is still true that each next particle
gives an
extra double log, so the answer should look as
\be
 \sigma_{gg\rightarrow (n-2)g} \approx \sigma_{gg\rightarrow gg}
[ \alpha_s N_c  C_n log^2(s/s_0)]^{n-4} \label{eq:sigman}
\ee
   Unfortunately, to get the  coefficient $C_n$ is not that
simple. That was done in ref.
\cite{Goldberg_Rosenfeld}
under a series of approximations, who have found the asymptotic coefficient
\be C_n \rightarrow const={1 \over 4\pi\sqrt{3}} \ee
   We have checked, that at least with the approximation made, the
limiting value of $C_n$ is correct. We have also calculated
in the same approximation the
non-leading log terms till constant, for n=6-10\footnote{
This was done by
SUNY student Wen-Chen Chang, whose help is acknowledged.}. In all cases the
next-to-leading coefficient is zero, so corrections go as
$1+const/log^2(s/s_0)+...$. Other coefficients are large  and they
generally have  varying signs. Therefore
 for the interesting value of $s/s_0 \sim 20$  all terms are
comparable and the leading-double-log approximation is unreliable.

Note however, that the leading log approximation indicate an interesting
problem by itself.
 Unlike in the deep inelastic or $e^+e^-$ annihilation case,
the ordering of virtuality and rapidity is not demanded, thus the corresponding
integrals are not 'nested'.
The number of 'strings' in the PT formula is growing factorially. As a result,
there is
 {\it no n-factorials in denominator}, and,  while summing up all
orders in gluon production, one gets a geometrical series instead of
exponential ones. If so, the multi-gluon
emission processes
limit applicability of
perturbation theory at {\it finite} energy! Non-perturbative methods (e.g.
based on instantons or other classical solutions) are
probably needed in order  to evaluate the
 high-energy
contribution of multi-gluon processes into the total hadronic cross section.

  Parke-Taylor matrix element and the leading log approximation predict also
   significantly different picture of the produced parton  distributions. This
is examplified in Fig.2, where a sample of transvers momentum $p_t$ and
rapidity distributions of secondary gluons are shown
for high-energy $\sqrt{s}=16 {\rm GeV}$
multigluon events ( $gg\rightarrow (n-2)g$ )
with a resolution $s_0= 0.3 {\rm GeV}^2 $.
Both distributions are normalized
for one particle
and the line types are respectively solid ( n=4 ),
dotted ( n=5 ), short dash ( n=6 ), long dash ( n=7 ) and dot-short dash
( n=8 ).
Going from n=4 (elastic process) to larger n
one can see that the particle distribution begin to build up very rapidly at
central rapidity.
When $n=5$, a soft emmision is quite obvious in addition to  two major
outgoing gluons. However when n is larger, all the outgoing particles
are distributed around $y=0$.
Its width is O(1), so the angular distribution is in fact nearly isotropic.
The $p_t$
spectrum becomes roughly exponential at larger n. Thus, something like
mini-fireball is produced in any multigluon event!

   These distribution are to be compared with 'soft gluon approximations',
predicting {\it flat} rapidity distribution $d\omega/\omega=dy$ and {\it
power like} $p_t$ spectra $d p^2_t/p^2_t$.

We have performed Monte-Carlo evaluation of the reaction rates for multi-gluon
production using tthe KS corrected Parke-Taylor matrix element, and our results
  for $n=6-8$ can be summarized as follows:
\be \alpha_6(\eta=1.1)= 0.924 \\
\alpha_7(\eta=1.1)= 0.201 \\
 \alpha_8(\eta=1.1)= 0.0278 \ee
   Their total contribution to gluon multiplication rate weighted by the
numberof particle being produced
\be \alpha_{multiplication} =\Sigma_{n=6-8} (n-4)\alpha_n = 2.56
\ee
is comparable to that of $gg \rightarrow ggg$ process, thus these higher-order
processes increase the gluon multiplication rate considerably.

In order to see more clearly which kinematical domain
is most relevant, we have plotted in Fig.3
the contributions of various invariant mass of the two initial gluons.
Generally for large n processes, the incoming energy of
the two gluons is required to be larger, in order to fulfill the
cut off constrains. We see that from $n=5-8$, the threshold
increases, by about $1.8T$ for adding a gluon.
The large-n  processes is therefore suppressed exponentially due to
the thermal distribution. We found that they roughly satisfy
\be
\alpha_{n}/\alpha_{n-1} = { 0.875^{n-4}\over n-4 },\ \ {\rm for}\ \ n\ge 5
\ee

It is somewhat amusing to compare this ratio to that given by the leading log
asymptotic expression discussed above
\be
\alpha_{n+1}/\alpha_{n}\approx \alpha_s N_c C_n \log^2 {s\over s_0} \approx 0.3
\ee
We find out that the geometric series result is not bad
for most relevant processes, although
the contribution of large-n processes drop faster than geometric series.
In any case, the problem mentioned above, divergence of total cross section,
is actually irrelevant for the thermal problem under consideration\footnote{It
may be a serious question while considering the very initial 'partonic'
before kinetic equilibration.}.

In the end of this section we want to add few remarks on the relation between
our approach and the PCM \cite{Geiger_Muller}. Not only we found that
the 'soft gluon' kinematics is not the dominant one, we seriously question
applicability of the
 Lipatov-Altarelli-Parisi  splitting function. Even in the soft gluon domain
there are significant
 disagreements with the Parke-Taylor formula we use.

  For one radiated
gluon, the leading order contribution is a double-log, not a single log:
it is because LAP splitting function contain a subtraction, including some
radiation inside the initial structure function. Whether such subtraction is
needed in the cascade, is questionable.

  For more than one radiated gluon, there is no agreement
even  in the coefficient of the leading logs. The reason is PCM
assume ordering of virtualities and rapidities, not demanded by the PT formula.
In other words, it looks like a particular
tree-like diagram (as followed in PCM using {\it probabilities}) actually
interfere
with other  tree diagrams, and those are included in the PT formula.

 These points
deserves further detailed studies.

\vskip 1cm
\centerline{\bf 5. Gluon chemical equilibration}
\vskip 1cm

In this section we study the gluon production inside expanding gluon plasma,
 solving equations for
gluon fugacity and temperature, changing with time. The procedure is
essentially similar to what was first done in \cite{Biro_etal}.

 We start at time $\tau_0$, at which
the momentum distribution is given by Eq.(\ref{eq:mom}), with some
initial fugacity
$\xi< 1$.
The {\it energy conservation} yields one integral of motion, which
for scaling on-e-dimensional expansion can be written as
follows
\be
\xi^{3/4} T^3\tau = \beta =const(\tau)  \label{eq:1}
\ee
where the newly introduced  parameter
$\beta$ is  constant in time and can be determined from the
initial condition.

As the next step, we  take into account the lowest order
relevant processes $ gg \leftrightarrow ggg$.
The  time derivative of the gluon number can be written
as
\be
{d (\rho\tau)\over d\tau } = ( \xi^2- \xi^3 ) \alpha_5 T^4\tau, \label{eq:2}
\ee
where $\rho$ is the number density;
 and $\alpha_5$ the reaction
rate coefficient shown in Eq.(\ref{eq:alpha45}).
The term ($\xi^2- \xi^3$) is due to the net effect of
particle production and annihilation and vanishes when the chemical
equilibration is reached.
With Eq.(\ref{eq:1}) and Eq.(\ref{eq:2}) together, one can
eliminate temperature and determine
the $\xi(\tau)$.  The problem is reduced to solving the following
differential equation
\be
{dy\over dx} =  (y^4- y^8){1\over x^{1/3} }\label{eq:3}
\ee
where
\be
y=\xi^{1/4}, \ \ \
x=\tau\beta^{1/2}(\alpha_5/1.6)^{3/2}  \label{eq:x}
\ee
are dimensionless variables.
At this point we notice immediately that the equilibration time scales
with the reaction rate coefficient by $\alpha_5^{3/2}$,
and with the ``initial entropy'' by $\beta^{1/2}$.
The faster the reaction, or the larger the initial entropy, the faster the
chemical equilibration.
The above equation can be solved analytically. The result is
\be
-{1\over 3y^3} +{1\over 2}\bigl[ {1\over 2}\log {1+y\over 1-y} + \tan^{-1}y
\bigr]
 = {3\over 2}x^{2/3} + const, \label{eq:solu}
\ee
with the constant being determined from the initial condition.

This approach can easily be extended to the  inclusion of
$ gg\leftrightarrow (n-2)g $.
Eq.(\ref{eq:2}) now becomes
\be
{d (\rho\tau)\over d\tau } =
\sum_{n=5}^\infty  (\xi^2  - \xi^{n-2} ) (n-4)\alpha_n
T^4\tau, \label{eq:4}
\ee
where the factor $(n-4)$ is the gluon number change in the processes.
After eliminating $T$ by using Eq.(\ref{eq:1}), the following differential
equation shows up
\be
{d\xi^{1/4} \over d\tau } =
\sum_{n=5}^\infty  (\xi  - \xi^{n-3} ) (n-4)\alpha_n /1.6
(\beta /\tau )^{1/3} \label{eq:expand}
\ee
This can be easily solved numerically.

(If one assumes the rates to form a geometric series
$
\alpha_n / \alpha_{n-1} =\chi=const(n),
$
one can obtain the analytical solution:
\be
(1-\chi)^2 [ -{1\over 3y^2} +{1-\chi\over 2}
( {1\over 2}\log{ 1+y\over 1-y} + \tan^{-1}y ) ]
= {3\over 2} x^{2/3} +const,
\ee
where $y,x$ are the same as defined before in Eq.(\ref{eq:x}).)

We have plotted the above results for  $T(\tau),\xi(\tau)$
in the upper and lower panels of Fig.4.
Three different initial conditions were chosen as examples:
$(\ T_i ({\rm GeV}), \ \xi_i \ ) = (\ 0.56,\  0.06\ ),
(\ 0.5, \ 0.25\ ),$ and $ (\ 0.5,\  0.5\ )$.
The dashed curves are the analytical results of $gg\rightarrow ggg$.
The dotted curves are also the analytical results for all n processes,
by assuming $\chi=0.2 $.
The solid curves are the solution to Eq.(\ref{eq:expand}) with n up to 8,
with $\alpha_n$ taken from the exact calculation.
We can see that the chemical equilibrium of gluons happen very fast.
For even very small initial $\xi_i$, the process $gg\leftrightarrow ggg$
alone can make the system to chemical equilibrium within about 3 fm/c.
The inclusion of more processes makes it faster, and
for large  $\xi_i >0.5$ the chemical equilibration happens within 1fm/c.

The time dependence of the temperature are in the top panel of Fig.4.
The types of curves are arranged in the same way as in the bottom panel.
Generally speaking the cooling of the non-equilibrium system
happens  faster than the chemical equilibrium system, simply
because some portion
of the thermal energy is contributed to creating new particles
during the expansion. However, as soon as equilibration is completed, $\xi=1$,
the cooling becomes adiabatic.
When the temperature drops to $T_c$
(where each line ends), our perturbative analysis breaks down.
 One can see, that the total lifetime of the plasma phase changes in this
 calculation from 5-6 fm/c for $\xi_i>1/2$ to only about 2 fm/c for
 $\xi_i\sim 1/20$.

  One can also observe from lower panel of Fig.4 that
if the initial density is too small, the expansion takes over and
equilibration does not take place before the critical temperature $T_c$ is
reached. Presumably this is what happens at lower energies (SPS etc).
  The existence of a 'sensitive window' around $\xi \sim 1/20$,
in which evolution may depend strongly on collision energy and
fluctuations  in the initial conditions, is a very
interesting phenomenon which can in principle be studied separately, on
event-per-event basis.

\vskip 1cm
\centerline{\bf 6. Conclusions and discussion}
\vskip 1cm

 To set a perspective, let us remind
the sequence of events in the 'hot glue' scenario:
\begin{itemize}
\item{} $0<\tau < \tau_{kin}\approx 0.3$ fm/c  : \hskip 1cm
entropy production, momenta equilibration
due to $gg \rightarrow gg$
The initial temperatures reached at this stage are estimated to be
 $T_i=400-500 MeV$  at RHIC and
 $T_i=600-900 MeV$  at LHC
\item{} $\tau_{kin} <\tau < \tau_{chem}$  :
 \hskip 1cm gluon chemical equilibration
due to  $gg \rightarrow ng$.
\item{} $\tau_{chem} <\tau < \tau_c=$ 3-5 fm/c : \hskip 1cm
adiabatic expansion of equilibrated  gluonic plasma,
terminated by the 'mixed phase' era, at which hadronic phase start to appear.
\end{itemize}

The  present  paper is devoted to studies of 'gluon multiplication'
period, and its conclusions are as follows.

(i) We have studied in details several processes leading to gluon
multiplication, using the exact result for $gg\rightarrow ggg$ and Parke-Taylor
formula for higher order processes.

(ii) The rates of the gluon multiplication processes
$gg \rightarrow (n-2)g$ form a convergent
series. They look like  $R_n/R_{n-1}\approx 0.875^{n-4}/(n-4) $ for up to
$n=8$. Total contribution
to multiplication rate of all processes producing more than one gluon
($n>5$) is
comparable to that of $gg\rightarrow ggg$ process.

 (iii) Including those rates in the equation for chemical equilibration in
longitudinally expanding plasma, we have found a solution for the fugacity and
temperature. The results depend critically on the initial conditions.
For large $\xi_i>.3-.5$ the chemical equilibration
time $\tau_{chem}$ can be as short as $\tau_{kin}$.

(iv) Even when the initial gluon number is  small, but not smaller
than $\xi\sim 1/10 $, the gluons still pile up
and the chemical equilibration is reached  within the
lifetime of the plasma.

(v) However, if  initial conditions happen to be
 around some
(so far very uncertain) small
critical value, large fluctuations on event-per-event basis may result,
affecting
even production of the
 total entropy (observed as total multiplicity or transverse energy)
 of the event.

(vi) The temperature decrease of the plasma is found to experience two
stages. The first one, during
 the chemical equilibration, correspond to relatively rapid cooling.
 The second one corresponds to adiabatic expansion. so $T\sim \tau^{-1/3}$
 all the way
towards the mixed phase, $T=T_c$.

   Finally, let us comment on the relation between
our approach and the
parton cascade model.
 The physics we consider, as well as the formulae used, are
clearly very different.
 Our study is focused on 'rapid multigluon production',
rather than on subsequent soft radiation by some cascade.
In spite of that, numerically
our conclusions are roughly consistent
with results obtained by K.Geiger \cite{Geiger_thermo}.

\vskip 1cm
\centerline{\bf 6. Acknowledgements}
\vskip 1cm
   Some very interesting discussions with K.Geiger and A.Makhlin
are acknowledged.
   This work is partly supported by the US DOE Grant. No. DE-FG02-88ER40388.

\vfill\eject
\newpage
\centerline { \bf Figure Captions }
\vskip  1cm

\noiondent{ Fig.1} The gluon elastic and the lowest order inelastic reaction
rate
coefficients ( defined in Eq.(\ref{eq:alpha}) )
 as a function of the medium cut-off ( $m_D =\eta T $).

\noindent{ Fig.2} The transverse momentum $p_T$ and rapidity $y$ distributions
of the final gluons in the c.m. frame of $gg\rightarrow (n-2)g$
multigluon processes.
The line types are respectively solid ( n=4 ),
dotted ( n=5 ), short dash ( n=6 ), long dash ( n=7 ) and dot-short dash
( n=8 ).
The c.m. energy is chosen to be $\sqrt s = 16 {\rm GeV}$
and the resolution of the gluons is $s_0 = 0.2 {\rm GeV}^2 $.

\noindent{ Fig.3} The rate spectra of the invariant mass of the two initial
gluons
that involve in the multi-gluon reaction
$gg\rightarrow (n-2)g$
in a fireball. The cut-off is from lattice data $\eta = 1.1$.

\noindent{ Fig.4} The time evolution of gluon fugacity $(\xi)$ and
temperature in an expanding gluon plasma,
for three sets of initial conditions
$(\ T_i ({\rm GeV}), \ \xi_i \ ) = (\ 0.56,\  0.06\ ),
(\ 0.5, \ 0.25\ ),$ and $ (\ 0.5,\  0.5\ )$.
The dash curves are the analytical results of $gg\rightarrow ggg$.
The dotted curves are also the analytical results for all n processes,
by assuming $\chi=0.2 $.
The solid curves are the solution to Eq.(\ref{eq:expand}) with n up to 8,
with $\alpha_n$ taken from the exact calculation.

%\bibliographystyle{try}
%\bibliography{ref}

\end{document}